\documentclass[prl,twocolumn,showpacs,preprintnumbers]{revtex4}
\usepackage{graphicx,bm,amsmath,amssymb,paralist,mathtools,dsfont}

\begin{document}

\newcommand{\avg}[1]{\langle #1 \rangle}
\renewcommand{\vec}[1]{\mathbf{#1}}
\renewcommand{\d}{\text{d}}
\newtagform{supplement}{(S}{)}

\title{Hot Brownian Motion}
  
\author{Daniel Rings}
\author{Romy Schachoff} 
\author{Markus Selmke}
\author{Frank Cichos}
\author{Klaus Kroy} \email{kroy@itp.uni-leipzig.de}

\affiliation{Institut f\"ur Theoretische Physik, Universit\"at
  Leipzig, Postfach 100920, 04009 Leipzig, Germany}

\affiliation{Institut f\"ur Experimentelle Physik I, Universit\"at
  Leipzig, Linnestra\ss e 5, 04103 Leipzig, Germany}

\date{\today}

\begin{abstract}
  We derive the generalized Markovian description for the
  non-equilibrium Brownian motion of a heated particle in a simple
  solvent with a temperature-dependent viscosity. Our analytical
  results for the generalized fluctuation-dissipation and
  Stokes-Einstein relations compare favorably with measurements of
  laser-heated gold nano-particles and provide a practical rational
  basis for emerging photothermal technologies.
\end{abstract}


\maketitle

Brownian motion is the erratic motion of suspended particles that are
large enough to admit some hydrodynamic coarse-graining, yet small
enough to exhibit substantial thermal fluctuations. Such mesoscopic
dynamics is ubiquitous in the micro- and nano-world, and in particular
in soft and biological matter
\cite{frey-kroy:2005,haw_middle-world:2006}.  Since their first
formulation more than a century ago, the laws of Brownian motion have
therefore found so many applications and generalizations in all
quantitative sciences that one may justly speak of a ``slow
revolution'' \cite{haw:2005}. In Langevin's popular formulation they
take the simple form of Newton's equation of motion for a particle of
mass $m$ and radius $R$ subject to a drag force $-\zeta_0\vec p/m$ and
a randomly fluctuating thermal force $\xi(t)$:
\begin{equation}
  \label{eq:langevin}
  \dot {\vec p} + \zeta_0 \vec p/m = \vec \xi   \qquad (t>0) \;.
\end{equation}
As a cumulative representation of a large number of chaotic molecular
collisions $\vec \xi$ is naturally idealized as a Gaussian random
variable. Its variance is tied to the Stokes friction coefficient
\begin{equation}
  \label{eq:stokes_law}
   \zeta_0=6\pi \eta_0 R 
\end{equation}
in a solvent of viscosity $\eta_0$ such as to guarantee
consistency of the averages $\avg{\dots}$ over force histories
$\xi(t)$ with Gibbs' canonical ensemble, namely
\begin{equation}
  \label{eq:FDT}
  \avg{\vec \xi(t)}=0\;, \qquad \avg{\xi_i(t)\xi_j(0)}=2k_BT_0\zeta_0  
  \delta_{ij} \delta(t) \;.
\end{equation}
This prescription implements the fluctuation-dissipation theorem for
the system comprising the Brownian particle and its solvent at
temperature $T_0$.  Strictly speaking, in view of how it deals with
long-ranged and long-lived correlations arising from conservation laws
governing the solvent hydrodynamics, this practical and commonplace
Markovian description applies only asymptotically for late times
\cite{mclennan:1988,keblinski-thomin:2006}.  Corresponding corrections
to Eqs.~(\ref{eq:stokes_law}-\ref{eq:FDT}) are accessible to modern
single-particle techniques and become most relevant in nano-structured
environments \cite{martin-etal:2006,jeney-etal:2008}.

Thanks to its prominent role in the ``middle world''
\cite{haw_middle-world:2006} between macro- and micro-cosmos, and its
experimental and theoretical controllability, Brownian motion has
become a ``drosophila'' for formulating and testing new (and sometimes
controversial) developments in equilibrium and non-equilibrium
statistical mechanics
\cite{blickle-etal:2006,blickle-prl:2007,howse-etal:2007,ritort:2008,dunkel-haenggi:2009,speer-eichhorn-reimann:2009,julicher-prost:2009}. In
this Letter, we introduce a non-equilibrium generalization that has so
far received little attention, namely the Brownian motion of a
particle maintained at an elevated temperature $T_{\rm p}>T_0$. From
its hypothetical sibling (``cool Brownian motion'', $T_{\rm p}<T_0$)
such ``hot Brownian motion'' (HBM) is distinguished by having obvious
realizations of major technological relevance such as nano-particles
suspended in water and diffusing in a laser focus. Due to a time-scale
separation between heat conduction and Brownian motion these particles
carry with them a radially symmetric hot halo easily detected with a
second laser. This provides the basis for promising photothermal
particle tracking \cite{bericiaud-etal:2004} and correlation
spectroscopy (``PhoCS'')
\cite{octeau-etal:2009,paulo-etal:2009,radunz-etal:2009} techniques
with a high potential of complementing corresponding fluorescence
techniques \cite{kim-heinze-schwille:2007} in numerous
applications. However, a photothermal measurement necessarily disturbs
the dynamics it aims to detect more severely than typical fluorescence
measurements, so that the development of an accurate theoretical
description of the Brownian motion of heated particles is a crucial
prerequisite for making the method competitive. This is not an
entirely straightforward task (as some might suggest \footnote{A
  common suggestion is to replace the ambient temperature $T_0$ and
  viscosity $\eta_0$ by $T_{\rm s}$ and $\eta(T_{\rm s})$,
  respectively.})  and requires an extension of the familiar theory,
as explained in the following. We arrive at simple analytical
generalizations of Eqs.~(\ref{eq:stokes_law}-\ref{eq:FDT}), which
should be sufficiently accurate for most practical applications.

For clarity, we restrict the following discussion to an idealized
situation: a hot spherical Brownian particle of radius $R$ at the
center of a co-moving coordinate system in a solvent with a
temperature-dependent viscosity $\eta(T)$ that attains the value
$\eta_0$ at the ambient temperature $T_0$ imposed at infinity.
Favorable conditions are assumed, such that potential complications
resulting from long-time tails \cite{jeney-etal:2008}, convection
\cite{katoshevski-etal:2001}, thermophoresis
\cite{dolinsky-elperin:2003}, etc.\ can be neglected. To avoid
confusion in comparisons with experimental data, we do however
distinguish the solvent temperature $T_{\rm s}$ at the hydrodynamic
boundary corresponding to the particle surface from the particle
temperature $T_{\rm p}$ itself, as these may differ substantially
\cite{merabia-etal:2009}. It is the temperature difference $\Delta
T\equiv T_{\rm s}-T_0$ that determines the heat flux responsible for
the non-equilibrium character of the problem. On relevant time scales,
the resulting temperature field around the particle follows from the
stationary heat equation, i.\,e.
\begin{equation}
  \label{eq:fourier}
  T(r)=T_0+ R \Delta T/r\;. 
\end{equation}
The task of finding appropriate generalizations of
Eqs.~(\ref{eq:stokes_law}-\ref{eq:FDT}) under these conditions is
split into two steps corresponding to the two force terms in
Eq.~(\ref{eq:langevin}), the damping and the driving force, or
friction and thermal noise, respectively. 

The first goal is mainly technical, namely to generalize
Eq.~(\ref{eq:stokes_law}) by solving
\begin{equation}
  \label{eq:stokes}
  \nabla \cdot \vec u =0 \;, \qquad   \nabla p = \nabla \cdot  \eta(r)
  [\nabla\vec u +(\nabla\vec u)^T] 
\end{equation}
for the stationary fluid velocity field $\vec u (\vec r)$ under the
usual no-slip boundary condition. The new feature compared to Stokes'
classical derivation is the radially varying viscosity $\eta(r)$
resulting from Eq.~(\ref{eq:fourier}). A numerically precise solution
of Eq.~(\ref{eq:stokes}) can be obtained with a differential shell
method \cite{rings-kroy:unpub} along the lines of similar work for
inhomogeneous elastic media \cite{levine-lubensky:2001}. However, for
our present purposes, as well as for practical applications, we wish
to find a generally applicable analytically tractable approximation.
We therefore resort to a toy model that evades the technical
difficulties related to the vector character of the fluid velocity but
retains the long-ranged nature of the hydrodynamic flow field. We
replace $\vec u(\vec r)$ by a fictitious diffusing scalar $u(\vec r)$
without direct physical significance, for which Eq.~(\ref{eq:stokes})
is readily solved analytically.  More explicitly,
Eq.~(\ref{eq:stokes}) reduces to $\nabla\cdot \eta(r)\nabla u(\vec
r)=0$ in the scalar model. A separation ansatz $u(\vec
r)=u_r(r)u_\vartheta(\vartheta)$ leads to the radial equation
\begin{equation}
  \label{eq:3}
  [\partial_r + 2/r+ (\partial_r\ln\eta)]\partial_r u_r= 0 
\end{equation}
solved by $\partial_r u_r \propto (\eta r^2)^{-1}$ for physically
reasonable functions $\eta(r)$. The quantities $u_r$ and
$\eta\partial_ru_r$ are now interpreted as the analogue of the
velocity of the particle and the hydrodynamic drag force per area,
respectively. The generalized effective friction coefficient
$\zeta_{\rm HBM}$ of hot Brownian motion is then estimated up to a
numerical factor as their ratio, disregarding the contribution from
the angular part. A comparison with Eq.~(\ref{eq:stokes_law}) in the
isothermal limiting case of constant viscosity $\eta(r)\equiv \eta_0$
helps to calibrate the model and fix the undetermined numerical
factor, which is then taken over to situations with radially varying
$\eta(r)$.  The accuracy of this procedure can be assessed and further
improved by a comparison with analytical and numerical results from
the mentioned differential shell method \cite{rings-kroy:unpub}. Some
technical details are provided in \cite{supplement} and the result is
summarized in Fig.~\ref{fig:eta_hbm}.

\begin{figure}
  \includegraphics[width=\linewidth]{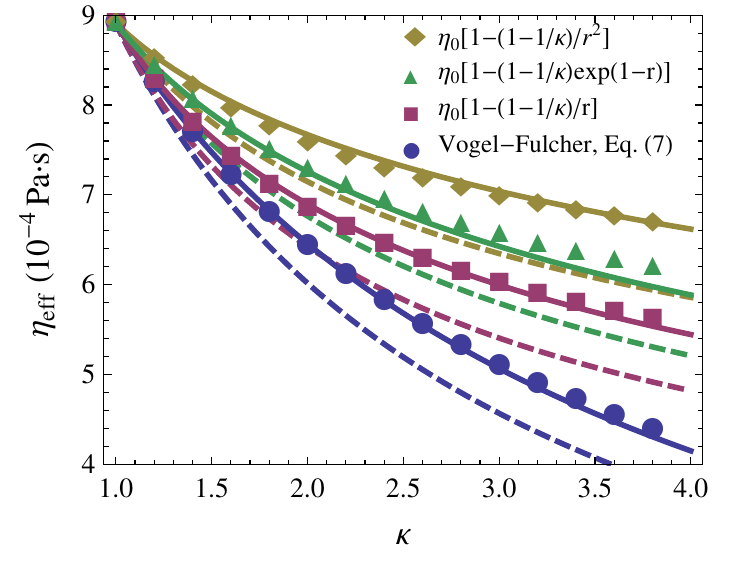}
  \caption{Comparison of analytical predictions of the scalar toy
    model to numerical results from the differential shell method
    (symbols) for exemplary long-ranged radial viscosity profiles
    $\eta(r)$ with $\kappa\equiv\eta(r\to\infty)/\eta(r=R)$ in a
    parameter regime of potential practical interest. While the
    simplest version of the model (dashed lines), which employs a
    constant calibration factor, exhibits noticeable systematic
    errors, the more elaborate version (solid line), corresponding to
    Eq.~(\ref{eq:etahbm}), should be sufficiently accurate for
    practical applications.}
  \label{fig:eta_hbm}
\end{figure}

An analytically tractable expression for the effective friction
coefficient $\zeta_{\rm HBM}$ as a function of temperature finally
results from a combination of the calibrated model with
Eq.~(\ref{eq:fourier}) and a phenomenological expression for the
temperature dependence of the solvent viscosity such as
\begin{equation}
\label{eq:VF}
\eta(T)=\eta_\infty \exp[A/(T-T_{\rm VF})] 
\end{equation}
(e.g.\ for water; but power-laws could be processed just as well).
The effective friction can be reinterpreted in terms of an
\emph{effective solvent viscosity} $\eta_{\rm HBM}\equiv \zeta_{\rm
  HBM}/6\pi R$ that replaces $\eta_0$ in Eq.~(\ref{eq:stokes_law})
under non-isothermal conditions. For reduced temperature increments
$\theta\equiv \Delta T/(T_{0}-T_{\rm VF})<1$ the result is well
approximated by its truncated Taylor series \cite{supplement} 
\begin{equation}
\label{eq:etahbm}
\begin{split}
  \frac{\eta_0}{\eta_{\rm HBM}} \approx 1 & + \frac{193}{486}
  \left[\ln \frac{\eta_0}{\eta_\infty}\right] \theta \\ &
  -\left[\frac{56}{243} \ln
    \frac{\eta_0}{\eta_\infty}-\frac{12563}{118098}
    \ln^2\frac{\eta_0}{\eta_\infty}\right] \theta^2 \;.
\end{split}
\end{equation}
This provides the wanted generalization of
Eqs.~(\ref{eq:langevin}-\ref{eq:stokes_law}).

To turn Eqs.~(\ref{eq:langevin}-\ref{eq:FDT}) into a fully predictive
Markov model of hot Brownian motion, the remaining task is to compute,
in the same spirit, an appropriate \emph{effective temperature}
$T_{\rm HBM}$ to replace $T_0$ in Eq.~(\ref{eq:FDT}).  In other words,
we aim at establishing a generalized \emph{non-equilibrium
  fluctuation-dissipation} relation for Brownian motion in a co-moving
radial temperature gradient. In analogy to the better understood
situation in globally isothermal non-equilibrium steady states
\cite{baiesi-maes-wynants:2009}, we expect to retrieve the
fluctuation-dissipation relation only after excluding the
``housekeeping heat'' from the entropy balance; i.e.\ the heat
constantly flowing from the particle to infinity to maintain the
temperature gradient.  All we have to consider is the minuscule
\emph{excess dissipation} associated with the damped motion of the
Brownian particle.  In this respect, it is crucial to appreciate the
long-range correlated character of the hydrodynamic flow, which
affects both dissipation and thermal fluctuations. It also helps in
setting up a systematic coarse-grained calculation by extending the
standard framework of fluctuating hydrodynamics
\cite{hauge-martin_lof:73} to moderate temperature gradients
\cite{rings-kroy:unpub}.

In simple terms, the process of Brownian motion can be rephrased as a
constant transformation of some thermal energy from the solvent into
an equal amount of kinetic energy for the Brownian particle and
\emph{vice versa}.  In a stationary situation the mutual energy
transfer must be balanced to obey the first law. More precisely, the
spatial integral over the local excess dissipation $\dot q(\vec r)$
--- i.\,e.\ the heat created (per unit of time) by the solvent flow at
position $\vec r$ in response to the movement of the Brownian particle
--- must on average match the rate of kinetic energy transfer $\dot
W_{\rm p}$ to the particle,
\begin{equation}
  \label{eq:first}
  \avg{\dot W_{\rm p}} = \int\!\d\vec r\; \avg{\dot q(\vec r)}\;.
\end{equation}
Moreover, to respect the second law, the motion must not cause a net
average entropy change, which was the origin of major reservations
against the modern interpretation of Brownian motion till the early
20th century. However, this only means that one has to make sure that
the integral over the local entropy flux to the solvent --- i.\,e.\ the
local dissipation rate $\dot q(\vec r)$ divided by the local solvent
temperature $T(\vec r)$ --- equals on average the entropy flux $\dot
S_{\rm p}=\dot W_{\rm p}/T_{\rm HBM}$ conferred to the Brownian
particle:
\begin{equation}
  \label{eq:second} 
   \int\!\d\vec r \; \frac{\avg{\dot q(\vec r)}}{T(\vec r)}
    = \frac1{T_{\rm HBM}}\int\!\d\vec r\; \avg{\dot q(\vec r) }\;.
\end{equation}
This then defines the wanted effective Brownian temperature $T_{\rm
  HBM}$, if the dissipation $\dot q(\vec r)$ is expressed in terms of
the local viscosity $\eta(\vec r)$ and $\nabla \vec u(\vec r)$. Within
our scalar model $\dot q(\vec r)=\eta(r)[\partial_ru_r(r)]^2/2$, hence
\begin{equation}
  \label{eq:toyT}
  T_{\rm HBM} =  \left.\int\!\d\vec r\; \eta (r)\avg{(\partial_r 
      u_r)^2} \! \right/ \!\!\! \int\!\d\vec r \; \frac{\eta
    (r)}{T(r)}  \avg{(\partial_r u_r)^2}\;.
\end{equation}
For the special case of a temperature-independent constant viscosity
$\eta_0$ this reduces to the simple explicit expression
\begin{equation}
  \label{eq:toyT0}
  T_{\rm HBM} = \Delta T/\ln (1+\Delta T/T_0)\;.
\end{equation}
The analytical expression generalizing this to the main case of
interest, a viscosity $\eta(r)$ that varies radially according to
Eqs.~(\ref{eq:fourier}) \& (\ref{eq:VF}), is given in
Ref.~\cite{supplement}. For small temperature increments $\Delta T\ll
T_0$ a practical approximation is
\begin{equation}
  \label{eq:Thbm}
  T_{\rm HBM} \approx T_0 +\Delta T/2 -\left[1-\ln(\eta_0/\eta_\infty)\right] \Delta T^2/(24T_0)\;.
\end{equation}

\begin{figure}
  \includegraphics[width=\linewidth]{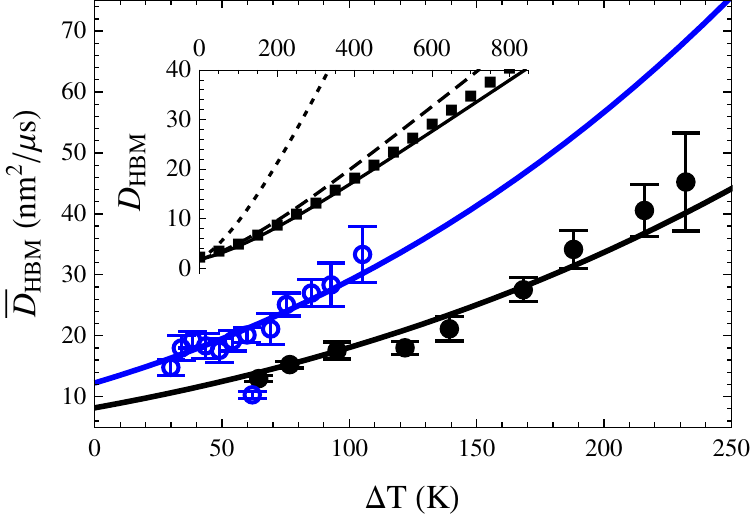}
  \caption{The effective diffusion coefficient $\bar D_{\rm
      HBM}(\Delta T)$ of hot gold nano-particles traversing a laser
    focus in water: experimental data (open/closed symbols for
    $R=40/60\;$nm) versus analytical predictions from the scalar model
    (solid lines); for solvent and focus parameters and the error bars
    see Ref.~\cite{supplement}. \mbox{\emph{Inset:}} $D_{\rm
      HBM}(\Delta T)$ according to numerical predictions from the
    differential shell method (squares), analytical solutions of the
    scalar model (corresponding to the lowest pair of curves in
    Fig.~\ref{fig:eta_hbm}), and the naive suggestion to identify the
    HBM parameters with the conditions at the particle surface
    (dotted); the agreement between the symbols and the solid line
    demonstrates the equivalence of Eqs.~(\ref{eq:second}) \&
    (\ref{eq:toyT}).}
  \label{fig:D_HBM_exp}
\end{figure}

A hot Brownian particle described by
Eqs.~(\ref{eq:stokes_law}-\ref{eq:FDT}) with $\eta_0$ and $T_0$
replaced by the corresponding effective quantities $\eta_{\rm HBM}$
and $T_{\rm HBM}$ from Eqs.~(\ref{eq:etahbm}) \& (\ref{eq:Thbm})
performs a random diffusive motion characterized by an effective
diffusion coefficient $D_{\rm HBM}$ obeying the generalized
Stokes--Einstein relation
\begin{equation}
  \label{eq:GSER}
   D_{\rm HBM} = \frac{k_BT_{\rm HBM}}{6\pi\eta_{\rm HBM}R} \;.
\end{equation}
This prediction is tested against the numerical differential shell
method in the inset of Fig.~\ref{fig:D_HBM_exp}. The good agreement
demonstrates the equivalence of Eqs.~(\ref{eq:second}) and
(\ref{eq:toyT}).

In order to test Eq.~(\ref{eq:GSER}) also experimentally, we used a
photothermal microscopy setup with gold nano-particles in water, as
described in Refs.~\cite{radunz-etal:2009,supplement}. Particles
passing through the common focal volume of a heating and a detection
laser beam leave a trace of photothermal bursts in the detector, which
encodes information about the diffusivity.  The spatially
inhomogeneous heating power in the laser focus implies, via
Eq.~(\ref{eq:GSER}), that the diffusion in the focus is
inhomogeneous. (A wider focus of the heating laser would avoid this
complication but is generally undesirable as one wants to minimize
sample irradiation.)  We therefore pursue a first-passage time
approach to determine the apparent effective diffusion coefficient
$\bar D_{\rm HBM}$ of \emph{inhomogeneous hot Brownian motion} from
the burst durations, which we identify with the transit times of the
particles passing through the focus volume \footnote{Note that the
  notion of apparent diffusion coefficients in inhomogeneous media
  is slightly ambiguous. Different generalizations of the homogeneous
  case pertain to different types of diffusivity measurements;
  S. Revathi and V. Balakrishnan, J. Phys. A: Math. Gen.  26, 5661
  (1993).}.  The time periods $\tau$ during which the photothermal
signal supersedes a fixed percentage of the maximum signal at a given
laser power are recorded for a large number of photothermal bursts.
The diffusion coefficient is then extracted from the exponential decay
of the obtained transit time distribution $P(\tau)$ at large $\tau$
\cite{ko:1997,nagar-pradhan:2003,supplement},
\begin{equation}
  \label{eq:effectiveD}
  \ln P(\tau \to \infty) \propto -\bar D_{\rm HBM}\tau  \;.
\end{equation}
Figure \ref{fig:D_HBM_exp} shows the result of such measurements for
various laser powers. The surface temperatures $T_{\rm s}=T_0+\Delta
T$ have been calculated from known quantities, namely the incident
laser intensity, the optical absorption coefficient of the particles,
and the heat conductivity of the solvent \cite{radunz-etal:2009}. Due
to our limited knowledge of the focus geometry, the factor of
proportionality in Eq.~(\ref{eq:effectiveD}) could not be determined
precisely, though. We therefore took the liberty to multiply each data
set by an overall factor to optimize the fit \cite{supplement}.  Yet,
the good agreement of the \emph{functional dependence} with the
prediction provides strong support for our analytical results, over a
considerable temperature range. At the same time, it establishes hot
Brownian motion as a robust and manageable tracer technique.

In summary, by introducing appropriate effective friction (viscosity)
and temperature parameters $\zeta_{\rm HBM}$ ($\eta_{\rm HBM}$) and
$T_{\rm HBM}$, for which we provided explicit analytical expressions
in Eqs.~(\ref{eq:etahbm}) and (\ref{eq:Thbm}), the convenient
Markovian description of Brownian motion in terms of
Eqs.~(\ref{eq:stokes_law}-\ref{eq:FDT}) could be extended to
non-equilibrium conditions, where the temperature of the Brownian
particle differs from that of the solvent. While
Eqs.~(\ref{eq:stokes_law}-\ref{eq:FDT}) are recovered in the
isothermal limit, the general predictions differ significantly from
what might have been guessed from simple rules of thumb and provide an
instructive illustration of the general dictum that hydrodynamic
boundary conditions should not be confused with the microscopic
conditions at the boundary
\addtocounter{footnote}{-1}\footnotemark[\value{footnote}]\addtocounter{footnote}{1}.
We sidestepped some technical difficulties of the corresponding
problem in fluctuating hydrodynamics by introducing an analytical toy
model that we calibrated with help of more elaborate analytical and
numerical calculations. Our analytical prediction for the effective
diffusion coefficient, based on the generalized Stokes--Einstein
relation in Eq.~(\ref{eq:GSER}), compares favorably with our
measurements of gold nano-particles depicted in
Fig.~\ref{fig:D_HBM_exp} and thus provides a convenient basis for
photothermal tracer techniques
\cite{bericiaud-etal:2004,radunz-etal:2009} with a high potential of
complementing corresponding fluorescence-based methods applied in many
fields from nano-technology to biology.

\begin{acknowledgments}
  We acknowledge helpful discussions with A. W\"urger, and financial
  support from the Deutsche Forschungsgemeinschaft (DFG) via FOR 877.
\end{acknowledgments}

\newpage

\usetagform{supplement}

\begin{widetext}
\renewcommand{\thefigure}{S\arabic{figure}}
\setcounter{figure}{0}
\setcounter{equation}{0}
\part*{Supplementary Material}

The supplementary material is organized as follows. In Section 1, we
revisit Stokes' problem of the viscous drag on a sphere for an
inhomogeneous viscosity $\eta(r)$ approximated by (i) a step
function and (ii) a staircase, from which we obtain numerically
exact solutions for $\eta_{\rm HBM}$ in the continuum limit. In
Section 2, we solve the scalar toy model. Comparison with (i)
suggests an improved calibration. Section 3 provides the complete
expression for the diffusion coefficient $D_{\rm HBM}$ and Section 4
the generalization $\bar D_{\rm HBM}$ for inhomogeneous hot Brownian
motion. Section 5 summarizes some phenomenological parameter
values. For complete derivations and a more comprehensive discussion
see \cite{Srings-kroy:unpub}.

\section{1. Shell and differential shell method for Stokes' problem}

Stokes' classical problem of finding the friction coefficient $\zeta$
of sphere of radius $R$ in a homogeneous Newtonian fluid of known
viscosity $\eta_0$ has a well-known solution: $\zeta=6\pi\eta_0 R$. As
stated in the main text, we wish to generalize this result to radially
varying viscosities $\eta(r)$. We consider
\begin{inparaenum}[(i)]
\item a step profile and
\item a staircase profile.
\end{inparaenum} In both cases the solution for the velocity $\vec{u}$
and pressure $p$ take on the form
\begin{equation}
  \label{eq:ansatz}
  \begin{aligned}
    u_r&=\left(a_0+\frac{a_1}{r}+a_2r^2+\frac{a_3}{r^3}\right)\cos\theta\\
    u_\theta&=-\left(a_0+\frac{a_1}{2r}+2a_2r^2-\frac{a_3}{2r^3}\right)\sin\theta\\
    p&=p_0+\left(\frac{a_1}{r^2}+10a_2r\right)\eta\cos\theta
  \end{aligned}
\end{equation}
in each of the spatially homogeneous regions.
\begin{enumerate}[(i)]
\item The step profile $\eta(r)=\eta_{\rm
    s}\left[1+(\kappa-1)\Theta(r-b)\right]$ jumping from $\eta_{\rm
    s}$ to $\eta_0\equiv\kappa\eta_{\rm s}$ at $r=b$. The coefficients
  $a_0\ldots a_3$ in the two domains $r<b$ and $r>b$ follow from the
  continuity conditions for the velocity and the stress at the jump
  discontinuity.
\item Similarly, the solution for the staircase profile is
  characterized by a set $(a_0, a_1, a_2, a_3)_j$ of coefficients and
  corresponding continuity conditions.  Along the lines of similar
  work for inhomogeneous elastic media \cite{Slevine-lubensky:2001}, we
  take the limit of an infinite staircase of infinitesimally thin
  shells, and the coefficients become radial functions $a_i(r)$ that
  can generally only be evaluated numerically. This \emph{differential
    shell method} yields the (inverse) effective friction coefficient
  \begin{equation}
    \label{eq:friction_coeff}
    \zeta^{-1}=\left.a_0(R)-a_1(R)R^{-1}+a_2(R)R^2+a_3(R)R^{-3}\right|_{F=1},
  \end{equation}
  evaluated at a reduced force of $F=1$. (The friction coefficient is
  defined by the force, \emph{viz.}\ the stress integrated over the
  particle surface, divided by the velocity relative to the fluid at
  infinity.)  Results for some exemplary viscosity profiles are shown
  as symbols in Fig.~1 of the main text.
\end{enumerate}

\section{2. Scalar Toy Model}
\label{sec:eff_visc}
To find a generally applicable analytically tractable approximation
for the friction coefficient $\zeta_{\rm HBM}$ we resort to a toy
model that evades the technical difficulties related to the vector
character of the fluid velocity but retains the long-ranged nature of
the hydrodynamic flow field. We replace $\vec u(\vec r)$ by a
fictitious diffusing scalar $u(\vec r)$ without direct physical
significance, for which Eq.~(5) reduces to $\nabla\cdot \eta(r)\nabla
u(\vec r)=0$. Hence, we seek a radially symmetric solution $u_r$ of
Eq.~(6), $[\partial_r + 2/r+ (\partial_r\ln\eta)]\partial_r u_r= 0
$. Integrating twice,
\begin{equation}
  \label{eq:scalar_visc_result}
  u_r(r) = K \int_r^\infty\frac{\d r'}{r'^2\eta(r')},\quad K=const.\;,
\end{equation}
which simplifies to $K/\eta_0 r$ for homogeneous viscosity
$\eta(r)\equiv \eta_0$, and can be expressed in the following closed
form for $T(r)=T_0+R\Delta T/r,\,\Delta T>0$ and the Vogel-Fulcher
temperature-dependence of the viscosity specified in Eq.~(7) of the
main text:
\begin{equation}
  \label{eq:u_stammfunktion}
  u_r(r) = -\frac{K\alpha}{\beta}\left[\frac{e^{-x}}{x}-\mathrm{Ei}(-x)\right]_\frac{\alpha}{1+\beta/r}^\alpha
\end{equation}
The abbreviations $\alpha\equiv B/T_0$ and $\beta\equiv R\Delta T/T_0$
have been used. The effective friction coefficient is
\begin{equation}
  \label{eq:eff_visc}
  \zeta_\text{HBM}=\frac{4\pi R^2\eta\partial_ru_r(R)}{u_r(R)}=\frac{4\pi K}{u_r(R)}\;.
\end{equation}
For homogeneous viscosity $\eta(r) \equiv \eta_0$ it degenerates to
$4\pi\eta_0 R$, indicating a mismatch by a factor of $3/2$ compared to
exact result. The simplest calibration of the scalar model consists in
correcting this constant factor such as to match the predictions in
the homogeneous case. In the inhomogeneous case, $\Delta T>0$, we
express Eq.~(S\ref{eq:eff_visc}) by the corresponding effective
viscosity $\eta_{\rm HBM}$. Including the mentioned factor of $3/2$
and making the temperature dependence (see Section 5) explicit, we
have (with the abbreviation $T^\ast\equiv T-T_\text{VF}$)
\begin{equation}
  \label{eq:eta0_etaHBM}
  \frac{\eta_0}{\eta_\text{HBM}}=\frac{e^{A/T^\ast} \left[A 
      \left(\text{Ei}\left(-\frac{A}{T^\ast+\Delta
            T}\right)- \text{Ei}\left(-\frac{A}{T^\ast}\right)\right)+(T^\ast+\Delta T)
      e^{-\frac{A}{T^\ast+\Delta T}}\right]-T^\ast}{\Delta T} \;.
\end{equation}
This result features as the lowest dashed line in Fig.~1 of the main
text; analogous calculations for different viscosity profiles
$\eta(r)$ provide the other dashed lines in the figure. Viscosity
profiles of the form $\eta(r)=\eta_0[1-(1-1/\kappa)(R/r)^n],\,n\in\mathds{N}$ result in
\begin{equation}
  \label{eq:eta_HBM_power_law}
  \frac{\eta_0}{\eta_\text{HBM}}=\,_2 F_1\left(1,\frac{1}{n};1+\frac{1}{n};\frac{\kappa-1}{\kappa}\right)\;,
\end{equation}
where $_2 F_1(a,b;c;z)$ denotes the hypergeometric function. For the
two cases $n=1,2$, shown in Fig.~1 of the main text,
$\eta_0/\eta_\mathrm{HBM}$ can be expressed as
\begin{equation}
  \label{eq:eta_HBMn=1}
  \frac{\eta_0}{\eta_\text{HBM}}=\frac{\kappa\ln\kappa}{\kappa-1}\;,
  \qquad (n=1)
\end{equation}
and
\begin{equation}
  \label{eq:eta_HBMn=2}
  \frac{\eta_0}{\eta_\text{HBM}}=\frac{\kappa\,\mathrm{arctanh}\sqrt\frac{\kappa-1}{\kappa}}{\sqrt{\kappa(\kappa-1)}}\;,
  \qquad (n=2) \;.
\end{equation}

For a more sophisticated calibration of the scalar model, we consider
again the step profile (i). The scalar model with the simple
calibration predicts the friction coefficient
\begin{equation}
\label{eq:scalar_step}
\zeta = \frac{b/R}{1+(b/R-1)\kappa}6\pi\eta_0R
\end{equation}
By $\kappa\equiv \eta_0/\eta_{\rm s}$, we denote the ratio of the
ambient viscosity and the solvent viscosity at the surface of the
Brownian particle, as before.  Note that the trivial limits $\kappa
\to 1$ and $b\to R$, and the limit $\kappa\to 0$ of a frozen surface
layer, are correctly obtained, whereas the joint limit $\kappa\to
\infty$ and $b\to R$, corresponding to a particle coated with an
infinitesimal superfluid layer, is ambiguous. To recover the correct
slip boundary condition ($\zeta=4\pi\eta R$) in this case, one has to
take this limit along the curve defined by $b/R=1+1/2\kappa$. If we
impose this constraint on Eq.~(S\ref{eq:scalar_step}), the calibration
factor to match it with the exact solution, which we obtain as
described in the previous section, \emph{viz.}\
Eq.~(S\ref{eq:friction_coeff}), is found as
\begin{equation}
  \label{eq:factor1}
   \frac{3 \left(80 \kappa ^4+80 \kappa ^3+60 \kappa ^2+20 \kappa
   +3\right)}{200 \kappa ^4+250 \kappa ^3+205 \kappa ^2+65
   \kappa +9}
\end{equation}
With this more elaborate calibration, the analytical predictions of
the scalar model (solid lines in Figures 1 and 2 of the main text)
practically coincide quite universally over a broad range of $\kappa$
with the numerical predictions obtained from the differential shell
method.

For moderate temperature increments $\Delta T\approx 0 \dots 150\,$K,
which are probably of greatest interest in practical applications, the
result can be further simplified by expanding
Eqs.~(S\ref{eq:eta0_etaHBM},S\ref{eq:factor1}) in a series in
$\theta\equiv\Delta T/(T_0-T_\mathrm{VF})$,
\begin{equation}
\label{eq:etahbm}
  \frac{\eta_0}{\eta_{\rm HBM}}  = 1 + \frac{193}{486} \left[\ln
    \frac{\eta_0}{\eta_\infty}\right] \theta -\left[\frac{56}{243} \ln
    \frac{\eta_0}{\eta_\infty}-\frac{12563}{118098}
    \ln^2\frac{\eta_0}{\eta_\infty}\right] \theta^2+O(\theta^3).
\end{equation}
Truncation after the second order yields Eq.~(8) of the main text. In
Fig.~\ref{fig:eta_eff_approx}, it is compared to the full expression
and to the result obtained with the simple calibration by the constant
factor $3/2$.
\begin{figure}
  \includegraphics[width=0.8\linewidth]{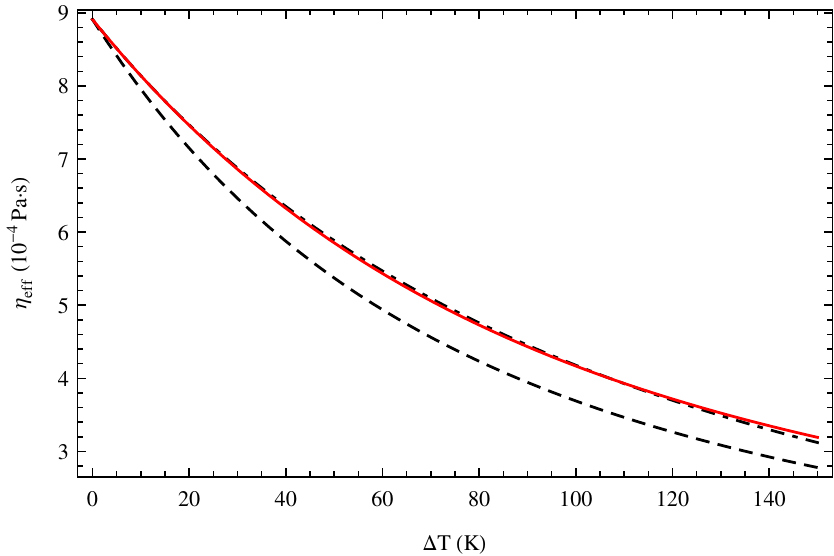}
  \caption{Temperature dependence of the effective viscosity according
    to the scalar model, Eq.~(S\ref{eq:eff_visc}): simplest
    calibration by a constant factor to match the isothermal limit
    (dashed); improved calibration with
    Eqs.~(S\ref{eq:eta0_etaHBM}-S\ref{eq:factor1}) (dash-dotted); the
    truncated Taylor series of Eq.(S\ref{eq:etahbm}) (solid).}
    \label{fig:eta_eff_approx}
\end{figure}

\section{3. Effective diffusion coefficient}

Inserting the solution $u(r)$ of the scalar model from
Eq.~(S\ref{eq:u_stammfunktion}) into Eq.~(11) of the main text yields
the effective temperature
\begin{equation}
  \label{eq:T_HBM_exact}
  T_\mathrm{HBM}=\frac{A e^{-\frac{X \left(T_0(\Delta T+T_0)-T_\mathrm{VF}^2\right)}{T_0 T_\mathrm{VF}}} \left[(T_0-T_\mathrm{VF}) e^{\frac{X (\Delta T+2 T_0-2 T_\mathrm{VF})}{T_0}} \left(\text{Ei}\bigl(-\frac{A}{T_0}\bigr)-\text{Ei}\bigl(\frac{A X}{T_0 Y}\bigr)\right)\bigr/T_0-e^{A/T_0}/X-e^{X-\frac{T_\mathrm{VF} X}{T_0}}/Y\right]}{e^{A Z}\left(\text{Ei}\bigl(\frac{A X}{T_0 Y}\bigr)-\text{Ei}\bigl(-\frac{A}{T_0}\bigr)\right)+\text{Ei}\bigl(-\frac{A}{T_\mathrm{VF}}\bigr)-\text{Ei}\bigl((\Delta T+T_0) X Z\bigr)},
\end{equation}
with the following abbreviations
\begin{equation}
  \label{eq:substitutions}
  X \equiv \frac{A}{\Delta T+T_0-T_\text{VF}},\quad
  Y\equiv \frac{A}{T_\text{VF}-T_0},\quad Z\equiv\frac{1}{T_0}-\frac{1}{T_\text{VF}}\;.
\end{equation}
Again, a simpler approximate expression is expected to suffice for
most practical purposes, in particular with regard to the various
minor contributions that were neglected altogether in our
approach. For moderate temperature increments $\Delta T\ll T_0$ a
Taylor expansion yields
\begin{equation}
  \label{eq:Thbm}
  \frac{T_{\rm HBM}}{T_0} = 1+\frac{\Delta
    T}{2T_0}+\left[\ln\Bigl(\frac{\eta_0}{\eta_\infty}\Bigr)-1\right]
  \frac{\Delta T^2}{12 T_0^2} + O\Bigl(\frac{\Delta T^3}{T_0^3}\Bigr) \;.
\end{equation}
The semi-phenomenological approximation quoted in Eq.~(13) of the main
text is obtained by dividing the second-order contribution by 2. As
illustrated by the comparison with the full expression,
Eq.~(S\ref{eq:T_HBM_exact}), in Fig.~\ref{fig:T_eff_approx}, the
deviations of the approximate expression Eq.~(13) from
Eq.~(S\ref{eq:T_HBM_exact}) are insignificant at the present overall
level of accuracy if $\Delta T$ is in the range $0\ldots 150$\,K. 

\begin{figure}
  \includegraphics[width=0.8\linewidth]{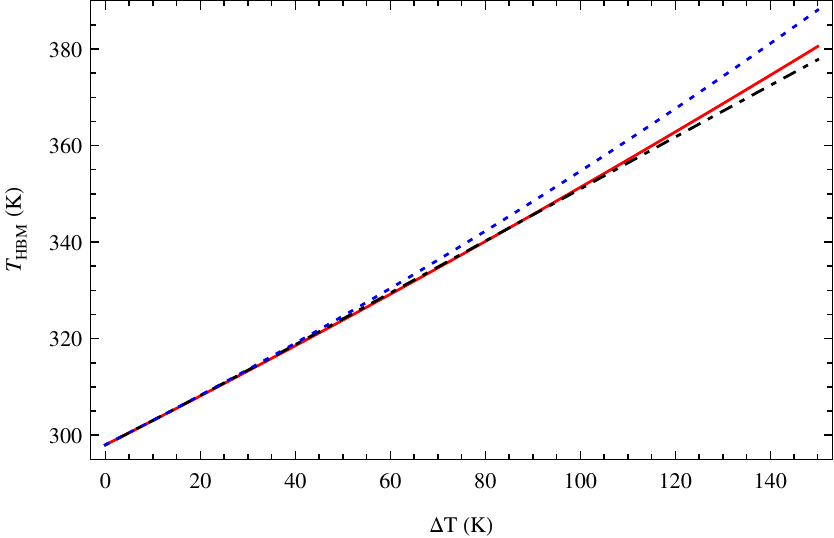}
  \caption{Dependence of various expressions for the effective
    Brownian temperature $T_\mathrm{HBM}$ on the temperature increment
    $\Delta T\equiv T_{\rm s}-T_0$: the calibrated scalar model given
    in Eq.~(S\ref{eq:T_HBM_exact}) (dot-dashed); the second order
    Taylor expansion Eq.~(S\ref{eq:Thbm}) (dotted, blue); and the
    semi-phenomenological approximation given in Eq.~(13), and
    displayed in Fig.~2, of the main text (solid, red). In contrast to
    the truncated Taylor series Eq.~(S\ref{eq:Thbm}), the
    semi-phenomenological expression has no significant errors over
    the practically relevant temperature range.}
    \label{fig:T_eff_approx}
\end{figure}

Inserting $T_\mathrm{HBM}$ from Eq.~(S\ref{eq:T_HBM_exact}) and the
effective viscosity $\eta_\mathrm{HBM}$ from
Eq.~(S\ref{eq:eta0_etaHBM}) into the generalized Stokes--Einstein
relation
$D_\mathrm{HBM}=k_\mathrm{B}T_\mathrm{HBM}/(6\pi\eta_\mathrm{HBM}R)$,
Eq.~(14) of the main text, we obtain the effective diffusion
coefficient
\begin{equation}
  \label{eq:D_HBM}
  \begin{split}
    D_\mathrm{HBM}&=\frac{\exp\left[-X (\Delta
        T+T_0+T_\text{VF})/T_\text{VF}\right]}{6 \pi \eta_\infty R T_0
      \Delta T \left(\text{Ei}\bigl(\frac{A (\Delta T+T_0) X}{T_0
            T_\text{VF} Y}\bigr)+e^{A
          Z}\left[\text{Ei}\bigl(-\frac{A}{T_0}\bigr)-
          \text{Ei}\bigl(\frac{A X}{T_0 Y}\bigr)\right]-\text{
          Ei}\bigl(-\frac{A}{T_\text{VF}}\bigr)\right)}\times\\
    &\negthickspace\negthickspace\negthickspace\left\{\left.A
        (T_0-T_\text{VF}) e^{\frac{X (\Delta T+2
            T_0-T_\text{VF})}{T_0}} \left[\text{Ei}\Bigl(\frac{A
              X}{T_0
              Y}\Bigr)-\text{Ei}\Bigl(-\frac{A}{T_0}\Bigr)\right]+T_0
        \left[\Delta T e^{\frac{A+T_\text{VF}
              X}{T_0}}-(T_0-T_\text{VF})
          \left(e^X-e^{\frac{A+T_\text{VF}
                X}{T_0}}\right)\right]\right\}\times\right.\\
    &\negthickspace\negthickspace\negthickspace\left\{-A e^X
      \text{Ei}(Y)+A e^X \text{Ei}(-X)+\Delta T-T_0
      e^{X+Y}+T_0+T_\text{VF} e^{X+Y}-T_\text{VF}\right\}\;,
  \end{split}
\end{equation}
with abbreviations as above. As illustrated in
Fig.~\ref{fig:D_HBM_approx}, a satisfying approximation is indeed again
obtained by use of the simple approximate expressions for
$T_\mathrm{HBM}$ and $\eta_\mathrm{HBM}$ from Eqs.~(8) \& (13) of the
main text, respectively.  This is how the curves in Fig.~2 of the main
text were generated.
 
\begin{figure}
  \includegraphics[width=0.8\linewidth]{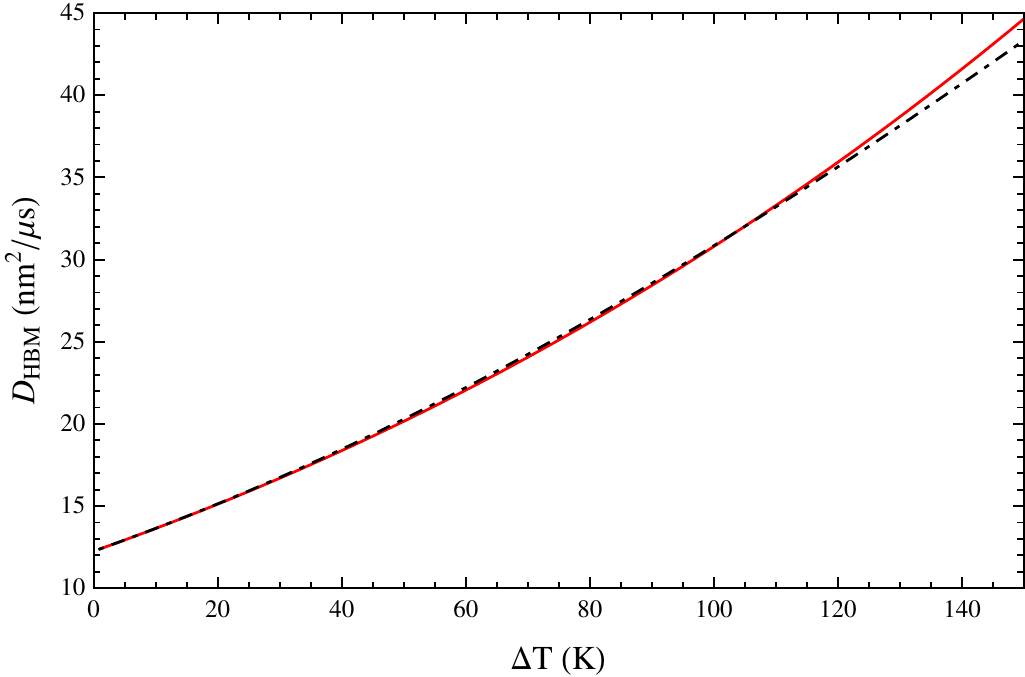}
  \caption{Temperature dependence of the effective diffusion
    coefficient $D_\mathrm{HBM}(\Delta T)$ from the scalar model, as
    given in Eq.~(S\ref{eq:D_HBM}) (dot-dashed), and in the
    approximation obtained by inserting the individual approximations
    of $T_\mathrm{HBM}$ and $\eta_\mathrm{HBM}$ from Eqs.~(8) \& (13)
    of the main text into the generalized Stokes-Einstein formula
    (solid, red).}
    \label{fig:D_HBM_approx}
\end{figure}

\section{4. Apparent diffusion coefficient for inhomogeneous
  heating power}

In practical applications diffraction usually limits the experimental
realization of a spatially uniform heating rate within the observation
volume. Therefore, the apparent diffusion coefficient deduced from the
transit time statistics of the particles passing through the focus
involves some implicit averaging over a spatially heterogeneous
$D_\mathrm{HBM}(\vec r)$. In the following, we derive a theoretical
expression for this average.

First, the local diffusion coefficient $D_\mathrm{HBM}(\vec r)$
follows immediately from the heating power density $I(\vec r)$ via
Eq.~(S\ref{eq:D_HBM}) by noting that $\Delta T(\vec
r)\propto I(\vec r)$ if the absorption coefficient of the particle is
temperature insensitive. Assuming a radially symmetric heating power
distribution in the focus, the appropriate averaging procedure is
similar to that for a particle released at the center of the focus,
which can be traced back to a standard first-passage-time problem
\cite{SRevathi1993,SZwanzig:2001}. The distribution $P(\vec r,t)$ of
escape times for the particle is obtained by solving the Smoluchowski
equation with an absorbing boundary condition at the boundary ${\cal
  B}$ of the focus volume,
\begin{equation}
  \label{eq:fokker_planck}
  \frac{\partial P}{\partial t}=\nabla\cdotp D(r)\nabla
  P
  ,\quad P(\vec r,0)=\delta(\vec r-\vec r_0),\quad
  P(\vec r,t)=0\text{ on } {\cal B}
\end{equation}
The spherically symmetric boundary value problem for the escape time
$\tau_{\rm p}(\vec r)=\tau_{\rm p}(r)$ of a particle starting at
position $\vec r$ is
\begin{equation}
  \label{eq:fpt_bvp}
  \tau_{\rm p}'(2/r+\partial_r)D+D\tau_{\rm p}''=-1 \quad \text{with} \quad \tau_{\rm p}(\vec r)=0\text{ on }
  {\cal B}\;,
\end{equation}
which has the general solution
\begin{equation}
  \label{eq:tau_solution}
  \tau_{\rm p}(r) = \int_\omega^r\left(\frac{c}{D(r')r'^2}-\frac{r'}{3D(r')}\right)\mathrm{d}r'\;,
\end{equation}
$\omega$ being the radius of the focus volume and $c$ a constant
of integration. For the \emph{escape problem} of a particle starting
in the center of the focus, $c=0$ is required by $\tau_{\rm p}(0)<\infty$. The
\emph{apparent} diffusion coefficient thus reads
\begin{equation}
  \label{eq:eff_diff_coeff}
  \bar D_\mathrm{HBM} = \omega^2/6\tau_{\rm p}(0)\;.
\end{equation}

For the related \emph{transit problem}, which is of interest for our
transit time analysis, the situation is slightly more complicated. The
most likely transit paths are only touching or barely entering the
focus, so that the transit time distribution $P_{\rm transit}(t)$
diverges at $t=0$. However, the characteristic transit time $\tau_{\rm
  t}$ may be extracted from the experimentally obtained transit time
distribution by fitting the asymptotic law $P_{\rm transit}(t\gg0)\sim
t^{-3/2}\exp[-t/\tau_{\rm t}]$ \cite{SKo:1997,Snagar-pradhan:2003}. The
stochastic errors inferred from the fits are displayed as error bars
in Fig.~2 of the main text. The apparent diffusion coefficient follows
from $\tau_{\rm t}$ as
\begin{equation}
  \label{eq:eff_diff_coeff_exp_sphere}
  \bar D_\mathrm{HBM} = \omega^2/\pi^2\tau_{\rm t}\;,
\end{equation}
for spherical focus geometry. In practice, the focus is usually more
elongated along the optical axis than transverse to it, so that it may
be better approximated by a cylinder, in which case
\begin{equation}
  \label{eq:eff_diff_coeff_exp_cylinder}
  \bar D_\mathrm{HBM} = \omega^2/\alpha_1^2\tau_{\rm t}\;,
\end{equation}
where $\alpha_1$ denotes the first zero of the Bessel function of the
first kind of order zero. The different numerical factors in the last
two expressions therefore provide a lower bound for the systematic
numerical uncertainties involved in the determination of the absolute
value of $\bar D_\mathrm{HBM}$.

A typical experiment yields a time series of photothermal bursts
\cite{Sradunz-etal:2009}. Their intensity is proportional to the local
power densities of the lasers used for heating the particle and
detecting the induced refractive index change, respectively. Here, an
uncertainty arises since the focus geometry cannot be controlled or
determined precisely in the diffusion experiment. A nominal lateral
focus size $\rho=300\,$nm has therefore been estimated by fitting a
Gaussian intensity profile $\propto\exp[-r^2/(2\rho^2)]$ to the
photothermal image of single immobilized gold nano-particles obtained
with the same setup. As the axial extension of the focus is usually
large compared to the lateral one (about $1\,\mu $m), the focal volume
is approximated by a cylindrical shape, corresponding to
Eq.~(\ref{eq:eff_diff_coeff_exp_cylinder}). Particles are identified
as ``in the focus'' if the signal intensity surpasses a certain
threshold set to a fixed percentage of the maximum signal of the whole
time trace of bursts. The threshold therefore defines the actual focus
size $\omega$ relevant for the burst width analysis, which stays
constant during the measurements of a given sample, due to scaling of
the threshold with the maximum signal. In Fig.~2 of the main text, we
use the value of $\omega$ as a freely adjustable overall fit parameter
to match the experimental data with the theoretical prediction for
$\bar D_\mathrm{HBM}$ and find $\omega\approx 250\,$nm for the
$R=60\,$nm particles and $\omega\approx 170\,$nm for the $R=40\,$nm
gold nano-particles. Note that $\omega$ may generally differ between
the measurements of different samples (\emph{viz}.\ particle sizes)
due to variations in the signal-to-noise ratio and the sample
geometry.

\section{5. Parameters for the solvent viscosity}
Parameter values used for all graphs throughout the main text and the supplementary
material:
\begin{align*}
  &\text{ambient temperature} & T_0&=298\,\mathrm{K}&\\
  &\text{dynamic viscosity of water} &\eta(T)&=\eta_\infty\exp[A/(T-T_\mathrm{VF})]\\
  &\quad\text{with} &\eta_\infty&=0.0298376\,\mathrm{mPa\cdotp s}\\
  & &A&=496.889\,\mathrm{K}\\
  & &T_\mathrm{VF}&=152.0\,\mathrm{K}
\end{align*}

\end{widetext}


\begin{thebibliography}{10}

\bibitem{frey-kroy:2005}
E. Frey and K. Kroy, Ann.\ Phys.\ (Leipzig) {\bf 14},  20  (2005).

\bibitem{haw_middle-world:2006}
M. Haw, {\em Middle World: The Restless Heart of Matter and Life} (Macmillan,
  New York, 2006).

\bibitem{haw:2005}
M. Haw, Physics World {\bf 18 (1)},  19  (2005).

\bibitem{mclennan:1988}
J.~A. McLennan, {\em Introduction to Non-Equilibrium Statistical Mechanics}
  (Prentice Hall, Englewood Cliffs, 1988).

\bibitem{keblinski-thomin:2006}
P. Keblinski and J. Thomin, Phys.\ Rev.\ E {\bf 73},  010502  (2006).

\bibitem{martin-etal:2006}
S. Martin, M. Reichert, H. Stark, and T. Gisler, Phys. Rev. Lett. {\bf 97},
  248301  (2006).

\bibitem{jeney-etal:2008}
S. Jeney {\it et~al.}, Phys.\ Rev.\ Lett. {\bf 100},  240604  (2008).

\bibitem{blickle-etal:2006}
V. Blickle {\it et~al.}, Phys.\ Rev.\ Lett. {\bf 96},  070603  (2006).

\bibitem{blickle-prl:2007}
V. Blickle {\it et~al.}, Phys.\ Rev.\ Lett. {\bf 98},  210601  (2007).

\bibitem{howse-etal:2007}
J.~R. Howse {\it et~al.}, Phys.\ Rev.\ Lett. {\bf 99},  048102  (2007).

\bibitem{ritort:2008}
F. Ritort, Adv.\ Chem.\ Phys. {\bf 137},  31  (2008).

\bibitem{dunkel-haenggi:2009}
J. Dunkel and P. H\"anggi, Phys.\ Rep. {\bf 471},  1   (2009).

\bibitem{speer-eichhorn-reimann:2009}
D. Speer, R. Eichhorn, and P. Reimann, Phys.\ Rev.\ Lett. {\bf 102},  124101
  (2009).

\bibitem{julicher-prost:2009}
F. J{\"u}licher and J. Prost, Eur.\ Phys.\ J.\ E {\bf 29},  27  (2009).

\bibitem{bericiaud-etal:2004}
S. Berciaud, L. Cognet, G.~A. Blab, and B. Lounis, Phys. Rev. Lett. {\bf 93},
  257402  (2004).

\bibitem{octeau-etal:2009}
V. Octeau {\it et~al.}, ACS Nano {\bf 3},  345  (2009).

\bibitem{paulo-etal:2009}
P.~M.~R. Paulo {\it et~al.}, J.\ Phys.\ Chem.\ C {\bf 113},  11451  (2009).

\bibitem{radunz-etal:2009}
R. Rad{\"u}nz, D. Rings, K. Kroy, and F. Cichos, J.\ Phys.\ Chem.\ A {\bf 113},
   1674  (2009).

\bibitem{kim-heinze-schwille:2007}
S.~A. Kim, K.~G. Heinze, and P. Schwille, Nature Meth. {\bf 4},  963  (2007).

\bibitem{katoshevski-etal:2001}
D. Katoshevski, B. Zhao, G. Ziskind, and E. Bar-Ziv, J.\ Aerosol Sci. {\bf 32},
   73   (2001).

\bibitem{dolinsky-elperin:2003}
Y. Dolinsky and T. Elperin, J.\ Appl.\ Phys. {\bf 93},  4321  (2003).

\bibitem{merabia-etal:2009}
S. Merabia {\it et~al.}, Proc.\ Natl.\ Acad.\ Sci. (USA) {\bf 106},  15113
  (2009).

\bibitem{rings-kroy:unpub}
D. Rings {\it et~al.}, unpublished.

\bibitem{levine-lubensky:2001}
A.~J. Levine and T.~C. Lubensky, Phys. Rev. E {\bf 65},  011501  (2001).

\bibitem{supplement}
See the appended Supplementary Material, pp.~5.

\bibitem{baiesi-maes-wynants:2009}
M. Baiesi, C. Maes, and B. Wynants, Phys.\ Rev.\ Lett. {\bf 103},  010602
  (2009).

\bibitem{hauge-martin_lof:73}
E.~H. Hauge and A. Martin-L\"of, J.\ Stat.\ Phys. {\bf 7},  259  (1973).

\bibitem{nagar-pradhan:2003}
A. Nagar and P. Pradhan, Physica A {\bf 320},  141   (2003).

\bibitem{ko:1997}
D.~S. Ko {\it et~al.}, Chem. Phys. Lett. {\bf 269},  54   (1997).

\end{thebibliography}

\begin{thebibliography}{1}

\bibitem[S1]{Srings-kroy:unpub}
D. Rings {\it et~al.}, unpublished.

\bibitem[S2]{Slevine-lubensky:2001}
A.~J. Levine and T.~C. Lubensky, Phys. Rev. E {\bf 65},  011501  (2001).

\bibitem[S3]{SRevathi1993}
S. Revathi and V. Balakrishnan, J. Phys. A {\bf 26},  5661  (1993).

\bibitem[S4]{SZwanzig:2001}
R. Zwanzig, {\em Nonequilibrium Statistical Mechanics} (Oxford University
  Press, USA, 2001).

\bibitem[S5]{SKo:1997}
D.~S. Ko {\it et~al.}, Chem. Phys. Lett. {\bf 269},  54   (1997).

\bibitem[S6]{Snagar-pradhan:2003}
A. Nagar and P. Pradhan, Physica A {\bf 320},  141   (2003).

\bibitem[S7]{Sradunz-etal:2009}
R. Rad{\"u}nz, D. Rings, K. Kroy, and F. Cichos, J.\ Phys.\ Chem.\ A {\bf 113},
   1674  (2009).

\end{thebibliography}
\end{document}